\documentclass[fleqn,usenatbib,useAMS]{mnras}
\usepackage{graphicx}	% Including figure files
\usepackage{amsmath}	% Advanced maths commands
\usepackage{amssymb}	% Extra maths symbols
\usepackage{multicol}        % Multi-column entries in tables
\usepackage{comment}
\usepackage{bm}		% Bold maths symbols, including upright Greek
\usepackage{pdflscape}	% Landscape pages
\usepackage{xcolor}

\def\lsim{\lower.5ex\hbox{$\; \buildrel < \over \sim \;$}}
\def\gsim{\lower.5ex\hbox{$\; \buildrel > \over \sim \;$}}
\def\be{\begin{equation}}
\def\bea{\begin{eqnarray}}
\def\eea{\end{eqnarray}}
\def\ee{\end{equation}}

\def\mp{m_{\rm p}}

\def\bc{\begin{center}}
\def\ec{\end{center}}

\def\etal{{\em et. al.}}
\def\ie{{\em i.e. }}

\def\rg{r_{\rm s}}
\def\rs{r_{\rm s}}

\def\vt{v_{\rm \small T}}

\def\ecal{\cal E}
\def\td{T_{\rm disc}}
\def\rin{r_{\rm in}}
\def\ro{r_{\rm o}}
\def\rd{r_{\rm d}}
\def\sigmt{\sigma_{\rm \small T}}
\include{netbib.sty}

\usepackage{graphicx}
\title[Winds from Accretion Discs]
{\textbf{Simulations} of radiation driven winds from Keplerian discs}
\author[Raychaudhuri et. al.]
{Sananda Raychaudhuri$^{1}$, Mukesh K. Vyas$^{2}$, Indranil Chattopadhyay$^{3}$ \thanks{Email:
sanandaraychaudhuri@gmail.com (SR);mukeshkvys@gmail.com (MKV);
		indra@aries.res.in (IC)}\\
 $^{1}$ Centre for Astroparticle Physics and Space Science, Department of Physics, Bose Institute, \\ 
   Block EN, Sector V, Salt Lake, Kolkata, India 700091. \\	
 $^{2}$  Bar Ilan University, Ramat Gan-5290002, Israel. \\  	
$^{3}$Aryabhatta Research Institute of Observational Sciences 
(ARIES), Manora Peak, Nainital-263002, India.\\
}
\begin{document}
\date{}
\maketitle
\label{firstpage}
\begin{abstract}
We study the ejection of winds from thin accretion discs around stellar mass black holes and the time evolution
of these winds in presence of radiation field
generated by the accretion disc. Winds are produced by radiation, thermal pressure and the centrifugal force 
of the disc. The winds are found to be mildly relativistic, with speeds
reaching up to terminal speeds $0.1$ for accretion rate $4$ in Eddington units. 
We show that the ejected matter gets its rotation by transporting
angular momentum from the disc to the wind. We also show that the radiation drag affects the accretion
disc winds in a very significant manner. Not only that the terminal speeds are reduced by an order of magnitude
due to radiation drag, but we also show that the non-linear effect of radiation drag,
can mitigate the formation of the winds from the matter
ejected by the accretion disc.
As radiation drag reduces the velocity of the wind, the mass outflow rate is reduced in its presence as well.
%\textbf{As radiation drag reduces the velocity of the wind, the mass outflow rate is reduced in its presence as well.}
%The mass outflow rate is much less ($\sim 10^{-4}$ Eddington rates) compared to the accretion rate and therefore the
%radiation field as well as the wind achieves quasi steady state.   
\end{abstract}


\twocolumn
\section[Introduction]{Introduction}
\label{sec:intro}
Outflows in the form of winds are commonly associated with various astrophysical sources like AGNs,
X-ray binaries, YSOs etc. In radio quiet AGNs, blue-shifted iron lines are frequently reported. 
This blue shift is believed to be generated from resonance absorption of Fe-xxv or Fe-xxvi by
propagating winds away from the source. The speeds of these winds are found to be relativistic and
may reach up to $0.4c$ \citep{CH02, CH03, Mrk06, DM05, C09, RJN09}. Similarly this blue shift is
observed in case of X-ray binaries as well \citep{ML07, TB16, GP12}.

These winds are observed in nearly half of these sources \citep{TF2010} indicating that the feature is quite general.
Further, their
short variability timescales ($\sim100$ks) suggest that the winds might be outflowing from the
central source from a region within 100 Schwarzschild radii ($\rs$). %With this information, we can immediately infer that
%the winds are generated from  the accretion disc \citep{C09}.
In X-ray binaries, the winds are observed in soft state only where the
observed spectra is mostly dominated by thermal emissions. In soft state, the accretion discs are well described by standard
thin disc
model, where the discs are optically thick but geometrically thin and emit thermally distributed radiation \citep{SS73}.

Though, theoretically these discs are highly stable
against most perturbations, but is susceptible to magneto-rotational instabilities \citep{BH91, SUZ2009, Yuan12}, %},
which apart
from providing an origin of shear viscosity, may also contribute to outflows.
Independent of such instabilities, magnetic field can remove energy and angular momentum from the accretion disc,
such that centrifugally driven outflow along the magnetic field is possible \citep{BP82}.

Not only magnetic field can drive winds, but
%\textbf{magnetic field can drive winds, but}%magnetically driven outflows, 
 winds can also be generated
by thermal and radiation pressure from the accretion discs \citep{BMS83}.
%{\bf 
It may be noted that outflows within sub-Eddington limit from optically thick discs were also studied
by performing MHD simulations \citep{OH2009,OH2011}.
\cite{LD2019} also studied the MHD outflows from the accretion discs in general relativistic limit. %}

As the winds are found to be traveling up to mildly relativistic speeds, they need driving agents. The radiation driving
of outflows (jets or winds) are studied by various authors through semi analytic works
\citep{F96, TF96,CC00a,CC00b,cc02, IC05,kcm14,
VKMC15, VC18, VC19} and radiation was shown to be an effective factor to accelerate jets and winds up to
relativistic speeds. Similarly simulations were also carried out to study the effects of radiation on outflows
\cite{PR97, PSD98, Y18, PR00, NO2017, PR03}. \cite{PR03b} and \cite{NO2017} studied line driven winds.
%\textbf{ and \cite{NO2017}} studied line driven winds.
However, the line force may not be that
effective when the temperature of the wind exceeds the ionization temperature significantly ($>10^5$ K). Hence the winds driven by
the radiation from inner
region of the accretion disc especially in microquasars, where the temperatures are hotter, one may need other mechanisms.
In that case, the radiation drives the winds
directly by depositing the momentum and/or energy. 
\citet{Y18} studied winds driven from hot corona
by radiation force of the underlying Keplerian disc (hereafter KD). It may be noted that, \citet{Y18}
did not consider the role of radiation drag although they maintained the optically thin condition 
through out in their simulation. Moreover, \citet{Y18} considered the outflow from a hot corona.
%{\bf \citet{Y18} studied winds driven from hot corona
%by radiation force of the underlying Keplerian disc (hereafter KD). It may be noted that, \citet{Y18}
%did not consider the role of radiation drag although they maintained the optically thin condition 
%through out in their simulation. Moreover, \citet{Y18} considered the outflow from a hot corona.}
In most of the previous attempts mentioned above, the radiation drag was not part of their analysis. %We will show that
%at low accretion rates, the radiation drag is sufficient to disrupt the winds to flow out of the computational domain.
%We will also calculate the amount of matter taken out from the disc through the winds.
We would like to investigate whether continuum emission of a thin accretion disc can radiatively drive matter to form a wind,
in presence of radiation drag.
Apart from this, we investigate how much angular momentum of the accretion disc is
transmitted to the winds above it.
We would also like to study the effect of radiation drag on the wind solution,
and will discuss the role of angular momentum removal in the winds due to radiation.
%The strength of the
%ejected matter is analyzed though the terminal speed of the  
%wind as it leave the computational domain. 
As this is an exploratory study, we intend to study how accretion rate affects these aspects of wind generation.
% As the wind takes away matter from the disc, it affects the rate of accreted matter leading to change of the standard spectra from the thin discs. Hence, along with the wind dynamics, we also give an estimate of the spectral shift due to this effect.

In section \ref{sec:assump}, we discuss the underlying assumptions, then we will show the set of governing equations and
the radiation field in section \ref{sec_equations}. Afterwards we describe the simulation set up, initial and boundary conditions
for the simulations, 
method of solving the equations, and numerical technique in section \ref{sec_numeri_approach}. We then proceed to results
(section \ref{sec_results}) and conclude the paper (section \ref{sec_conclusions}) with the significance of the analysis.
%\textbf{In section \ref{sec:assump}, we discuss the underlying assumptions, then we will show the set of governing equations and
%the radiation field in section \ref{sec_equations}. Afterwards we describe the simulation set up, initial and boundary conditions
%for the simulations, 
%method of solving the equations, and numerical technique in section \ref{sec_numeri_approach}. We then proceed to results
%(section \ref{sec_results}) and conclude the paper (section \ref{sec_conclusions}) with the significance of the analysis.}

\section{Assumptions}
\label{sec:assump}
We perform hydrodynamic simulation in cylindrical coordinate system $r,\phi,z$.
Axisymmetry in the system is assumed. In our simulation, the source of the wind is KD around a $10 M_\odot$ black hole,
which occupies the equatorial plane and winds
are launched from the KD in the $r-z$ plane.
The radiation field above the KD interacts with the out-flowing wind through Thomson scattering, and drives the wind by
depositing momentum onto the matter. We
restrict ourselves to non relativistic regime and hence, while calculating the radiation field above the disc, relativistic
transformations are ignored. However, to take care of strong gravity near the central source, we have assumed the Paczy\'nski \&
Wiita potential \citep{PW} that mimics the general relativistic effects.
In this paper, all ten independent components of the moments of radiation field are calculated, hence the
effect of radiation drag is also incorporated. In this paper the distances are scaled and shown in Schwarzschild units.

\section{Governing equations}
\label{sec_equations}
%We need to solve the set of hydrodynamic conservation equations, after including the required source terms. We will write the Momentum equations and then discuss about the source terms in this section.
%\begin{equation}
The equations of motion for a fluid in the radiation-hydrodynamic regime
\citep[correct up to first order in $v_i$, see ][for details]{mm84,kfm98}
with
density $\rho$, pressure $p$, propagating with velocity components
$v_i\equiv (v_r, v_\phi{~\rm and~} v_z)$, are given by 

\begin{equation}
\frac{\partial{\rho}}{\partial t} + \frac{1}{r}\frac{\partial (r\rho v_r)}{\partial r} + \frac{\partial{(\rho v_z)}}{\partial z} = 0
\label{eq.cont}
\end{equation}

\begin{eqnarray}
 \label{eq.momentum_r}
\frac{\partial{(\rho v_r)}}{\partial t}&+& \frac{1}{r}\frac{\partial (r\rho v_r^2)}{\partial r} + 
\frac{\partial p}{\partial r} + \frac{\partial{(\rho v_r v_z)}}{\partial z} \\ \nonumber
&=& \rho v_{\phi}^2/r +
 \rho f_{{\rm g},r} + \frac{\rho k}{c} {\cal F}_r %( F^r - E v_r - v_r P^{rr} - v_{\phi} P^{r \phi} - v_z P^{rz})  
\end{eqnarray}
%\begin{multline}
\begin{eqnarray}
 \label{eq.momentum_phi}
\frac{\partial{(\rho v_{\phi})}}{\partial t} &+& \frac{1}{r}\frac{\partial (r\rho v_r v_{\phi})}{\partial r} +
\frac{\partial{(\rho v_z v_{\phi})}}{\partial z} \\ \nonumber
&=& - \frac{\rho v_r v_{\phi}}{r} + \frac{\rho k}{c}{\cal F}_\phi %( F^{\phi} - E v_{\phi} - v_r P^{\phi r} - 
%v_{\phi} P^{ \phi \phi} - v_z P^{\phi z})
 \end{eqnarray}
%\end{multline}
%\begin{multline}
\begin{eqnarray}
\label{eq.momentum_z}
\frac{\partial{(\rho v_z)}}{\partial t} &+& \frac{1}{r}\frac{\partial (r\rho v_r v_z)}{\partial r} 
+ \frac{\partial{(\rho v_z^2+P)}}{\partial z} \\ \nonumber
&=&
\rho f_{{\rm g},z} + \frac{\rho k}{c}{\cal F}_z %( F^z - E v_z - v_r P^{z r} - v_z P^{zz} - v_{\phi}P^{z \phi}) 
\end{eqnarray}
%\end{multline}
%\end{equation}
\begin{equation}
\frac{\partial \ecal}{\partial t} + \frac{1}{r}\frac{\partial [r({\ecal}+p)v_r]}{\partial r} +
\frac{\partial{[({\ecal}+p)v_z}]}{\partial z} = -\rho\left(\frac{k}{c}v_i{\cal F}_i+v_if_{{\rm g},i}\right)
\label{eq.energy}
\end{equation}
In the above equations, ${\ecal}=\rho v^2/2+e$ is the energy density of the fluid and $e=p/(\Gamma -1)$ is the thermal energy
density, where $\Gamma=5/3$ is the adiabatic index of the fluid. The local sound speed is defined
as $c_s=(\Gamma p/\rho)^{1/2}$. 
%\textbf{The local sound speed is defined
%as $c_s=(\Gamma p/\rho)^{1/2}$.}
The fluid is being driven by the radiation
field of the underlying thin accretion disc. The components of radiation terms are:
\begin{equation}
{\cal F}_i= F_i - E v_i - v_r P_{ir} - v_{\phi} P_{i \phi} - v_z P_{iz},
\label{radcomp.eq}
\end{equation}
where $i\equiv (r,\phi,z)$. Various moments of the radiation field are
$E$, $F_i$s and $P_{ij}$s (here, $i,j \rightarrow r,\phi, z$), which
are basically the radiation energy density, various components of radiation flux and various components of radiation pressure,
respectively. The scattering opacity is $k=\sigmt/\mp$ with $\sigmt$ being Thomson
scattering cross section and $\mp$ is the proton mass. Moreover,
 $f_{{\rm g}r}$ and $f_{{\rm g}z}$ are $r$ and $z$ components of the gravitational force and are given by 
\begin{equation}
f_{{\rm g}z} = \frac{GM}{\rg^2} \frac{z}{R(R/\rg - 1)^2},
\end{equation}
\begin{equation}
f_{{\rm g}r} = \frac{GM}{\rg^2} \frac{r}{R(R/\rg - 1)^2},
\end{equation}
where $G$ and $M$ are the universal constant of gravity and the mass of the black hole, respectively. The Schwarzschild radius
of black hole defined as $\rg=2GM/c^2$. All the lengths mentioned in the paper are in terms of $\rg$ and we will refer
to them in dimensionless form afterwards. %\textbf{All the lengths mentioned in the paper are in terms of $\rg$ and we will refer
%to them in dimensionless form afterwards.}
Further, $R$ is the radial distance from the centre of the black hole defined as
\begin{equation}
R = \sqrt{r^2 + z^2}
\end{equation}
In the R.H.S %\textbf{R.H.S}
of the components of momentum equations (\ref{eq.momentum_r}, \ref{eq.momentum_phi} and \ref{eq.momentum_z}), the
components of radiative flux accelerates, while other velocity dependent terms with $E$ and $P_{ij}$, have negative
sign and therefore decelerates the flow. These are called radiation-drag terms and they show that radiation can also reduce the
momentum of the flow. As the radiation drag depends upon various components of the fluid velocity, it becomes effective as the flow
speed increases.
%\textbf{As the radiation drag depends upon various components of the fluid velocity, it becomes effective as the flow
%speed increases}.
The
radiative acceleration and deceleration depend upon the relative strengths of various radiative moments and the components of
flow speeds,
therefore, the effect of radiation on the outflow can behave in a very nonlinear manner. We will show in section
\ref{sec_results} that radiation acceleration drives the
winds to the infinity. However, consideration of radiative drag term reduces the outflow speed, to the extent that it can even
disrupt the ejected winds. Below we discuss the accretion disc and the radiative moments computed from its radiation field.

\subsection{Accretion disc properties}
An accretion disc around a black hole, on one hand, supplies matter to the black hole, on the other hand also supplies matter
flowing out as outflow. In the present case, the outflow is driven by the disc radiation.
Since the KD is defined on the equatorial plane
so the dynamical coordinates of the KD is represented by $R_{\rm d}\equiv (\rd,\phi, 0)$. From the mass conservation equation, we have the expression of accretion rate to be
\begin{equation}
\label{eqn:masscon}
\dot{M} = 2\pi \rd \rho v_{r{\rm K}} (2H),
\end{equation}
where, $H$ is the height of the disc from equatorial plane.
$v_{r{\rm K}}$ is the radial inflow speed due to accretion. The KD rotation velocity is \citep{PW,kfm98},
\begin{equation}
v_{\rm K} = \sqrt{\frac{GM\rd}{(\rd-\rg)^2}}
\label{vfi.eq}
\end{equation}
In KD $v_{\rm K}>>v_{r{\rm K}}$, and 
the radial velocity distribution is given by \citep{kfm98} 
\begin{equation}
v_{r{\rm K}} = 3.1 \times 10^6 \alpha^{\frac{4}{5}} \dot{m}^{\frac{2}{5}} 
             m^{-\frac{1}{5}}x^{-\frac{2}{5}} 
              \left(1-\sqrt{\frac{3}{x}}\right)^{-\frac{3}{5}},
\label{radvel.eq}
\end{equation}
where $x=\rd/\rg$ and $\alpha$ is the viscosity parameter.
%Here $\alpha$ is viscosity coefficient that is taken to be $0.1$ in this paper. 
Now, the distribution of the equatorial density along $\rd$ %\textbf{along $\rd$}
is obtained to be \citep{SS73}
\begin{equation}
\rho = 4.423 \times 10^4 m^{-1} \dot{m} x^{-1} v_{r{\rm K}}^{-1}
\label{dens.eq}
\end{equation}
The distribution of density along $z$ is $\tilde{\rho}=\rho e^{-{(z/\rg)}^2}$ \citep{SS73}.
It may be noted that at high accretion rates the inner part of the disc may become radiation pressure
dominated and in such cases the disc thickness is controlled by vertical radiative pressure rather than by gas pressure.
The density profile would change and instability might set in. However, we assume that since
radiation pressure is driving winds from the inner region, so the density profile of the disc may not depart significantly from
equation \ref{dens.eq}.
%\textbf{It may be noted that at high accretion rates the inner part of the disc may become radiation pressure
%dominated and in such cases the disc thickness is controlled by vertical radiative pressure rather than by gas pressure.
%The density profile would change and instability might set in. However, we assume that since
%radiation pressure is driving winds from the inner region, so the density profile of the disc may not depart significantly from
%equation \ref{dens.eq}.}
For a KD, viscosity is required for angular momentum transport in a manner such that the matter occupies subsequent
Keplerian orbits.
Viscosity heats up the matter, the dissipated heat is locally radiated as blackbody emission at each radius of the disc.
Assuming the surface temperature ($\td$) as the temperature of each annulus, its radial distribution is given by
\begin{equation}
\sigma {\td}^4 = \frac{3GM\dot{M}}{8\pi \rd^3}\left(1-\sqrt{\frac{\rin}{\rd}}\right),
\label{temp1.eq}
\end{equation}
where $\sigma$ is Stefan-Boltzmann’s constant, $\rin =3$ (in units of $\rs$)  is the inner radius of the disc, and the disc extends
up to an outer boundary $\ro=512$.
Expressing accretion rate and mass of the black hole in units of Eddington accretion rate, the previous equation becomes:
\begin{equation}
\td = 4.35 \times 10^7 \dot{m}^\frac{1}{4} m^{-\frac{1}{4}} x^{-\frac{3}{4}}\left(1-\sqrt{\frac{3}{x}}\right)^{1/4},
\label{temp2.eq}
\end{equation}
where, ${\dot m}={\dot M}/{\dot M}_{\rm Edd}$ and $M=m~M_\odot$, moreover, the Eddington accretion rate is ${\dot M}_{\rm Edd}=
1.44\times 10^{17}m$ (gm s$^{-1}$) and $M_\odot=2\times10^{33}$gm.

%Also, or the $z$ component of the flow in the disc domain, we put $v_z=0$.
%{\bf It is worth mentioning that the stability of the disc equations at super-Eddington luminosities leading to wind's
%ejection is a crude approximation. This approximation can be justified with the fact that the outflowing matter should be a
%very small fraction of the infalling matter and the disc doesn't evolve with time, retaining its Keplerian nature. We will analyze
%the outflow rates in the section \ref{sec_outflow} and will compare it with accretion rates.}
\subsection{Radiation field above a thin accretion disc}
%\begin{figure*}
%\centering
%\includegraphics[width=11cm]{Rflux10.eps}
%\includegraphics[width=10cm]{fig1.eps}
%\vspace{10mm}
%\caption{Contours of radiative moments computed at each point in $r-z$ plane around the black hole which resides at the origin and the disc on the equatorial plane. Energy density (a); radiative flux terms, %$F_r$, $F_z$ and $F_\phi$ (b)-(d) respectively; and the $F_r$, $F_\phi$ and $F_z$ (b)-(d), respectively; and the
%components of radiation pressure tensor, %$P_{rr},P_{r\phi},P_{rz},P_{zz},P_{\phi z},P_{\phi\phi}$
%$P_{rr},P_{r\phi},P_{rz},P_{\phi\phi},P_{\phi z}, P_{zz}$
%from (e) to (j), respectively.Only the inner $51.2 \times 51.2\rg$ region is shown.}
%\label{lab_rad_field}
%\end{figure*}

\begin{figure*}
%\floatbox[{\capbeside\thisfloatsetup{capbesideposition={left,center},capbesidewidth=4cm}}]{figure}[\FBwidth]
{\caption{Contours of radiative moments computed at each point in $r-z$ plane around the black hole which resides at the origin
and the disc on the equatorial plane. Radiation energy density (a); radiative flux terms, $F_r$, $F_z$ and $F_\phi$ (b)-(d)
respectively; and the components of radiation pressure tensor, $P_{rr}$, $P_{r\phi}$, $P_{rz}$, $P_{\phi\phi}$, $P_{\phi z}$, $P_{zz}$
from (e) to (j), respectively. Only the inner $51.2\rg \times 51.2\rg$ region is shown.}
\label{lab_rad_field}}
{\includegraphics[height=21cm,width=10cm]{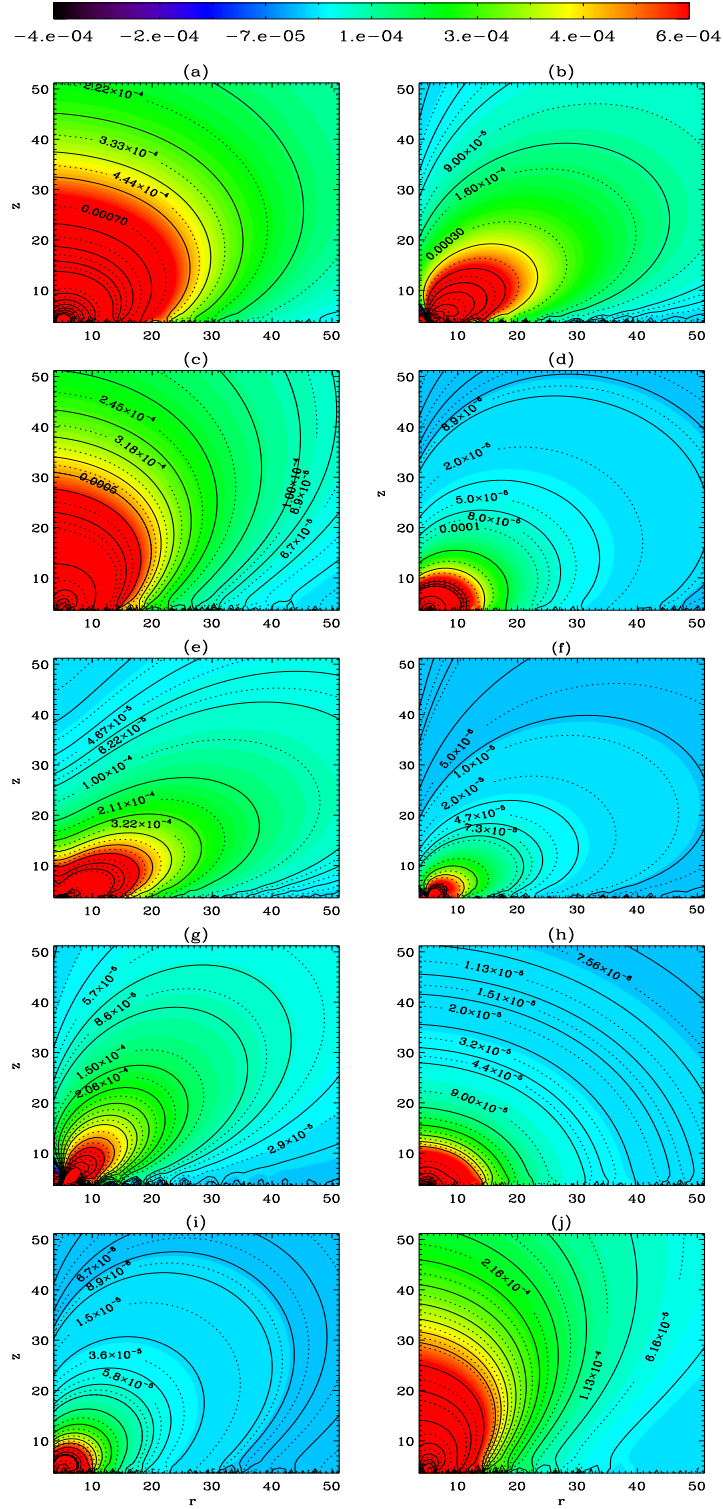}}
\end{figure*}
%\begin{figure}[!htb]
%  \captionsetup{singlelinecheck = false, format= hang, justification=raggedright, font=footnotesize, labelsep=space}
%  \centering
%  \begin{measuredfigure} % \begin{measuredfigure}
%    \includegraphics{fig1.eps}
%    \caption{Chuck Jones – Pepe Le Pew in action}
%  \end{measuredfigure}
%  \label{PlP}
%\end{figure}

In the following we present the expression of various radiative moments. 
For the convenience of representation, we define the radiative moments in the following forms,
%\begin{eqnarray}
$$
\frac{kE}{c}  =  {E}_{0}{\varepsilon};~~~~\frac{kF_i}{c} = F_{0}  f_i~~~{\rm and}~~~\frac{kP_{ij}}{c} = P_{0} p_{ij} 
$$

%\end{eqnarray}
{with, 
$$
{E}_{0}=F_{0}=P_{0}= \frac{3GM_B\dot{M}_K \sigma_T}{8 \pi^2 r_s^3 m_p c}
%\frac{4.32 \times 10^{17} \sigma_T \dot{m}}{32 \pi^2 m G M_{\odot}}
$$

%while the ambient is in hydrostatic equillibrium at t=0.
The dimensionless radiation energy density ($\varepsilon$), the three components of radiative flux ($f_i$) , as well as the six
components of pressure tensor ($p_{ij}$) are given by \citet{IC05}; 
\begin{eqnarray}
{\varepsilon} = {\int}^{\ro}_{\rin}
{\int}^{2 \pi}_0\frac{z(r^{-2}_{\rm d}-{\sqrt {3}}r^{-5/2}_{\rm d})d{\phi}^{\prime}
}{(r^2+z^2+r^2_{\rm d}-2rr_{\rm d}{\rm cos}{\phi}^{\prime})^{3/2}(1-v_il_i)^4
}dr_{\rm d}% \\ \nonumber
 %+ {\int}^{2{\pi}}_{{\phi}_f}\frac{z(x^{-2}_K-{\sqrt {3}}x^{-5/2}_K)d{\phi}^{\prime}
%}{(r^2+z^2+x^2_K-2rx_K{\rm cos}{\phi}^{\prime})^{3/2}
%}]dx_K
%\\ \nonumber
\label{eq:radeng}
\end{eqnarray}
%Three components of radiative flux term :
\begin{eqnarray}
%\begin{multline}
 f_i  =  {\int}^{\ro}_{\rin}
{\int}^{2 \pi}_0
\frac{z(r^{-2}_{\rm d}-{\sqrt {3}}r^{-5/2}_{\rm d}){\hskip 0.1cm}
l_id{\phi}^{\prime}}
{(r^2+z^2+r^2_{\rm d}-2rr_{\rm d}{\rm cos}{\phi}^{\prime})^{3/2}(1-v_il_i)^4
} dr_{\rm d} 
%\end{multline}
\label{eq:radflux}
\end{eqnarray}

%Six components of the pressure tensor are represented as follows:
\begin{eqnarray}
p_{ij} =  {\int}^{\ro}_{\rin} {\int}^{2 \pi}_0
\frac{z(r^{-2}_{\rm d}-{\sqrt {3}}r^{-5/2}_{\rm d}){\hskip 0.1cm}
l_i{\hskip 0.1cm}l_jd{\phi}^{\prime}}
{(r^2+z^2+r^2_{\rm d}-2rr_{\rm d}{\rm cos}{\phi}^{\prime})^{3/2}(1-v_il_i)^4
}dr_{\rm d},
%\\ \nonumber
\label{eq:radpres}
\end{eqnarray}
where  $l_i$s are the direction cosines from the disc to the field point.

%\textbf{
Since the accretion
disc is not a static radiator, but the disc matter is in motion, therefore the radiation field
is Doppler beamed by this disc motion. It can be shown that the frequency integrated radiation intensity measured
by the comoving observer ($I_0$) has the following transformation relation with that measured by an inertial observer ($I$)
\citep{kfm98}
\begin{equation}
 \frac{I_0}{I}=\gamma^4(1-v_il_i)^4 \approx (1-v_il_i)^4
 \label{eq:dopfac}
\end{equation}
The Lorentz factor $\gamma \approx 1$ for KD. This factor appears in the expression
of the moment equations and affects the radiation field. In particular the disc
motion along $\phi$ direction generates non-zero $f_\phi$ and also various components of $p_{i \phi}$.%}
The
coordinates of the thin Keplerian disc are ($r_{\rm d},\phi^\prime$)
and the integration limits of accretion disc are $\rin$= $3$ and
$\ro= 512$. 
%\section{Numerical method and Simulation setup}

We plot the dimensionless radiation moments \ie $E$ (Figure \ref{lab_rad_field}a), radiative
fluxes $F_r$ (Figure \ref{lab_rad_field}b), $F_z$ (Figure \ref{lab_rad_field}c), $F_\phi$ (Figure \ref{lab_rad_field}d) and the 6 independent components of
radiative pressure $P_{rr}$ (Figure \ref{lab_rad_field}e), $P_{r\phi}$ (Figure \ref{lab_rad_field}f), $P_{rz}$ (Figure \ref{lab_rad_field}g),
$P_{\phi \phi}$ (Figure \ref{lab_rad_field}h), $P_{\phi z}$ (Figure \ref{lab_rad_field}i) and $P_{zz}$ (Figure \ref{lab_rad_field}j).
The moments are plotted in $r-z$ plane. Each panel zooms the inner $51.2~\times ~51.2$ in order to resolve the contours of the
radiative moments. The radiative moments are distinctly anisotropic, especially close to the black hole. Since the KD only extends
up to $3$, and the KD flux maximizes at $r\sim 4$, so the radiative moments maximizes at around $4-5$. $E$ or the radiative energy density is
by far the most dominant of all the moments. Close to the axis $F_r\approx F_\phi \approx 0$, while $F_z$ is very important. In general,
$|F_z| \gsim |F_r|$ and dominates $F_\phi$. In addition, $P_{\phi \phi}$ is quite strong, hence the azimuthal velocity gained by the wind
due to $F_\phi$ will also be reduced due to the radiation drag along $\phi$ direction. Moreover, none of the components of the radiative
pressure is greater than all the radiative flux components. This would confine the effect of radiative drag.
This augurs well for the wind, so that it can be driven away from the KD, but would not be spread by a very large angle
due to the gain in angular momentum from the radiation field. We have plotted the radiative moments in a region very close to the
horizon ($\leq 51.2$), and in that region, the radiation field is from an extended source (KD), and therefore the moments show a 
complicated space dependence. At large distances, the space dependence of radiation field follows $\sim R^{-2}$, although not in the computational domain
we have chosen.
 
\section{Numerical approach}
\label{sec_numeri_approach}
\subsection{The numerical scheme and simulation set up}
The hydrodynamic equations (\ref{eq.cont}-\ref{eq.energy}) are solved in this paper using Total Variation Diminishing (TVD) scheme,
introduced and developed by \cite{AH83}. The scheme (or, the modified version of it) is applicable to hydrodynamic problems and has
been used extensively in relevant astrophysical
applications \citep{DR93,rbol95,rjf95,lrc11,IC12,lckhr16}. 
TVD scheme is an Eulerian, second order accurate, nonlinear, finite difference
scheme, which accurately captures shock.
%{\bf
The temporal and spatial evolution of the conserved quantities $\rho$,  $\rho v_i$,  and  ${\cal E}$
is computed using approximate Roe type Riemann solver to solve the differential equations, followed by application of a
non-oscillatory first order accurate scheme to the modified flux functions to achieve second order accuracy 
\citep[see,][]{ROE81, DR93, AH83}. %}
 %We can express the
Equations of motion (\ref{eq.cont}-\ref{eq.energy}) are similar to those solved in \citet{IC12}. In \citet{IC12} the galactic
outflow was powered by the radiation
from the galactic disc, while being decelerated by the gravity of the galactic disc, the halo and the bulge matter. In contrast, in
this
paper, the accretion disc outflow is powered by the radiative fluxes and the centrifugal force from the KD, and is decelerated by
the radiative
drag terms as well as, the gravity of the central black hole. 
To solve the equations of motion (\ref{eq.cont}-\ref{eq.energy}), we considered the TVD scheme
\citep[see,][for details]{IC12} for the resolution $512 \times 512$. 
%Following cell centred scheme, We define the initial parameters at the centre of each cell and solve Riemann problem at respective cell boundaries.
A schematic representation of the computational arrangement is presented in %\textbf{Figure}
Figure \ref{lab_grid}, which marks the ghost
cells where
the boundary
conditions are implemented and also the computational domain. %{\bf
We employed continuous boundary condition at $z=0$ boundary, %},
and outflow boundary condition at the outer $r$ and $z$ boundaries (i. e., no inflow but continuous if ${\bf v}>0$).
%\textbf{
At
$r=0$, or
the axis of symmetry, %}, 
reflection boundary condition has been employed. The type of boundary conditions employed, are also
mentioned in Figure (\ref{lab_grid}). We simulate a region of $512$ from the black hole, each in $r$ and $z$ direction, therefore 
the dimension of each cell is equivalent to $1$. The gravity of the black hole
is described by Paczy\'nski \& Wiita potential \citep{PW}. In order to avoid the coordinate singularity on the horizon, the black
hole
is covered by a sink region of radius $3$ around the origin, which do not affect the physics since the inner edge of KD is
$3$.

The KD is on the equatorial plane, ranging from $3-512$, the density, pressure and the components of velocity distribution
given by
equations (\ref{eqn:masscon} -\ref{temp2.eq}) are maintained in a region described by $r\rightarrow 3$---$512$ and
$z\rightarrow 0$---$3$. We supply the dynamical variables of the KD at every time step within a height of $3$
above the equatorial plane.
Therefore, KD acts as a boundary condition and is not dynamically sustained, so one may say its a quasi-KD.
The moments of the radiation field are computed in a region outside the KD.
We compute the outflowing winds due to the action
of the radiation field of a KD. 

The speed of light in vacuum
$c$ is the unit of velocity in the code, the unit of length is $\ro=512$.
Since the KD flux used in the code is for $m=10$ so the unit of time is $5.12\times 10^{-2}$ s. The reference %\textbf{
density is
$\rho_{\rm ref}=10^{-5}$ gm cm$^{-3}$. %}. \textbf{
The ambient medium or the computational domain is kept initially ($t=0$) tenuous
enough w.r.t the accretion disc with constant parameters. The ambient density is considered to be as low as a factor of $10^{-8}$
and pressure is $10^{-9}$ in code units, so the outflow from the disc is not suppressed artificially by the initial distribution
of right above the disc. %}
% The ambient density is taken to be $\rho_{amb}=10^{-8}$ while the ambient pressure is kept at $p_{amb}=10^{-9}$ in code units \ie these are scaled by $\rho_{\rm ref}$.}

\begin{figure}
\centering 
\includegraphics[width=80mm]{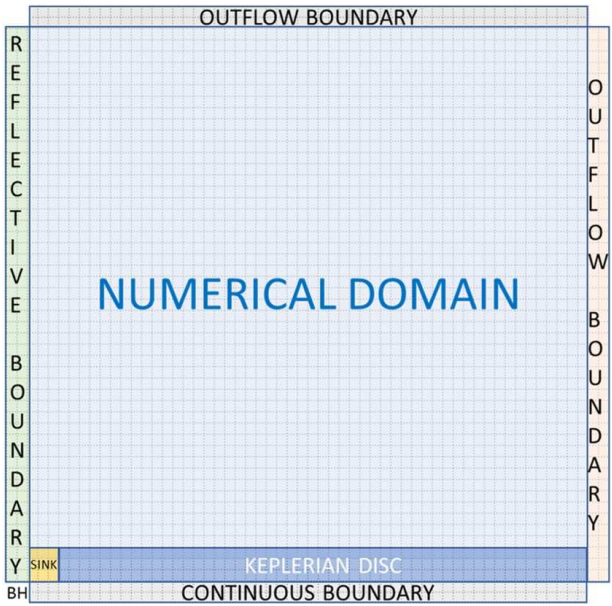}
\caption{Uniform grids in computational domain, with two ghost cells in each boundary. Respective boundary conditions are
mentioned accordingly. Grids drawn are not to the scale.}
\label{lab_grid}
\end{figure}
 %With the help of a TVD code developed to solve the problem, we compute the time evolution of these
%quantities and derive other related physical parameters.% which depend upon them.

%\begin{figure*}[]
%\centering
%\includegraphics[width=180mm]{fig3.eps}
%\caption{Contours of Density $log_{10}(\rho)$ over plotted with respective net velocity vector arrows.
%These profiles are for $\dot{m} = 3$. Panels correspond to the snapshots at time $t=2, 6, 62, 72, 82$ and $92$ from (a) to (f).} 
%\label{lab_den_md3}
%\end{figure*}

\begin{figure*}
  \begin{minipage}[c]{\textwidth}
    \begin{center}
        \includegraphics[width=180mm]{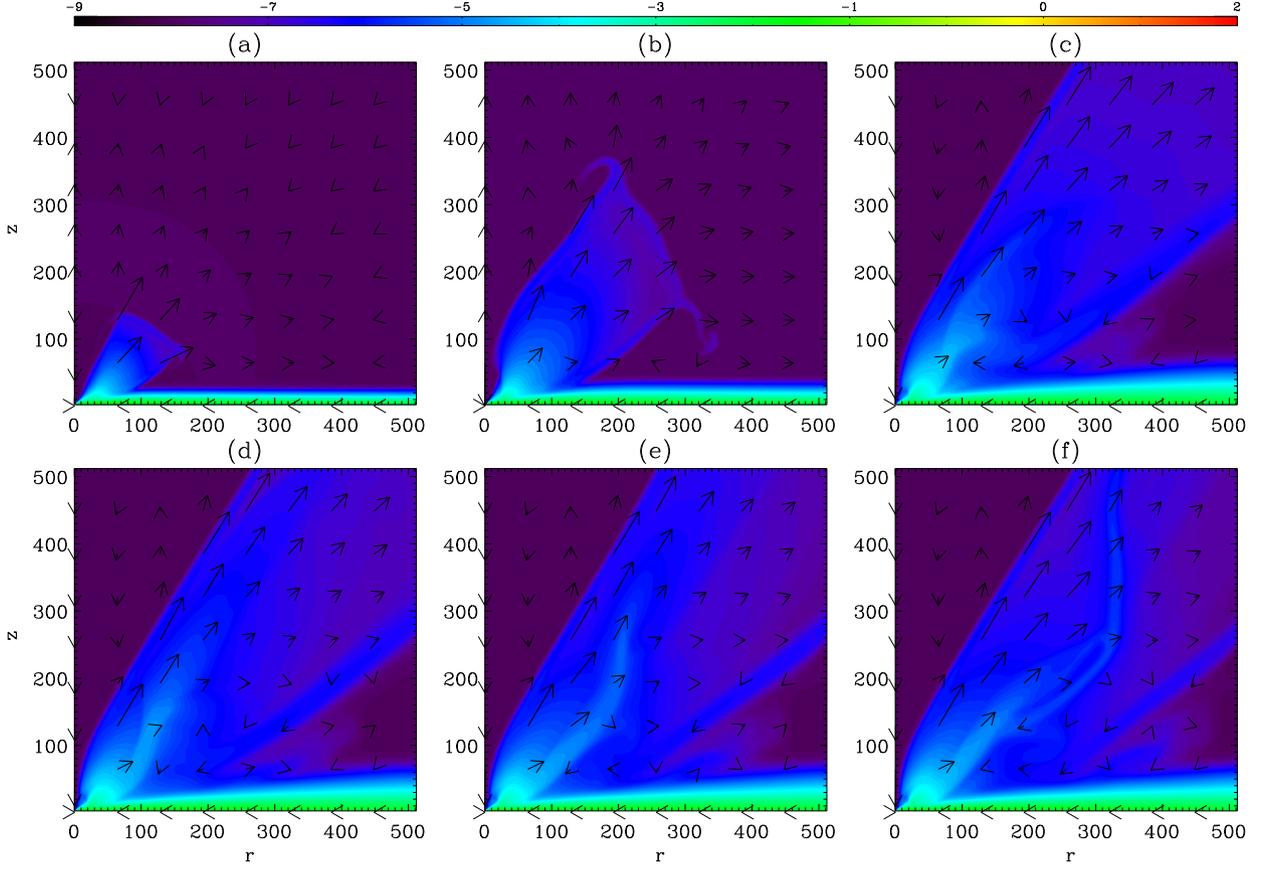} 
        \caption{Contours of Density $log_{10}(\rho)$ over plotted with respective net velocity vector arrows.
These profiles are for $\dot{m} = 3$. Panels correspond to the snapshots at run time $t=2, 6, 62, 72, 82$ and $92$ from (a) to (f).} 
	\label{lab_den_md3}
    \end{center}
  \end{minipage}
\end{figure*}

\section{Results}
\label{sec_results}

%\adjustimage{width=180mm,center,caption={some caption},label={somelabel},nofloat=figure,vspace=\bigskipamount}{fig3.eps}

%\begin{figure}
%\centering
%\adjustimage{width=180mm,center}{fig3.eps}
%\label{lab_den_md3}
%\caption{$lab_den_md3$}
%\end{figure}

\subsection{Wind propagation above the disc: density and velocity evolution}
In Figure \ref{lab_den_md3}(a)-(f), we overplot velocity vectors ($v_{\rm p} \equiv \sqrt{[v_r^2+v_z^2]}$) on the density contours of
radiatively driven
winds from a KD, for ${\dot m}=3$ and at different time steps $t=2$ (a), $t=6$ (b), $t=62$ (c), $t=72$ (d), $t=82$ (e), and $t=92$
(f). Arrows represent velocity vectors in $r-z$ plane, where the magnitude of the velocity ($v_p$) is proportional to the
length of the arrows.
All densities in this paper are scaled %\textbf{to} 
to $\rho_{\rm ref}$. 
The KD is hotter and denser near the inner edge, and the radiative flux maximizes at around $4$. Therefore,
both the thermal gradient force and the radiative force drive matter in the form of wind from the inner parts of the disc.
Very little matter is ejected from the region $\rd > 100$, even if the simulation is run for a longer time.

As the wind emerges from the inner regions of the disc, the general direction of motion is away from the axis of symmetry
[Figure \ref{lab_den_md3}(b)].
However, at a later time, a part of the wind moves towards the axis of symmetry [Figure \ref{lab_den_md3}(d)], but the wind again
moves away from the axis. The entire wind-fan oscillates as a whole, somewhat dancing like a flame in a breeze.
All the matter that is being ejected, does not flow out, but a tiny fraction of it falls back, and hits the wind base, which causes
a perturbation propagating along the wind. Moreover, $F_r$ near the wind base is directed towards the axis, but higher up it is
directed away from the axis. %\textbf{
The inner radius of KD is $r=3$, so there is no source of radiation for $r < 3$.
Hence, close to the axis of symmetry and just above
the disc, $r$ component of the radiative flux points inward \ie $F_r<0$. %}. 
The centrifugal force is always directed away from the axis. $F_\phi$ which is weaker than the fluxes
in the other two directions, will spin up the wind, but stronger pressure components boosts the drag in the $\phi$ direction.
Additionally the radiative force along $z$ powers the wind upwards.
And finally gravity attracts every part of the wind towards the black hole. All these factors together interact with the ejected
matter and generates a wind which originates from the inner region of the KD, but fans out in the $r-z$ plane.
%Strong $F_r$ towards the axis, acts against the outwards centrifugal 
%force and the wind is collimated partially. As the wind progresses upward, $F_r$ becomes weaker and the collimation vanishes and the 
%matter again recedes away from the axis. However, all the matter ejected do not flow out, a part of 
%the wind matter falls back towards the accretion disc and it hits the bottom of the wind base  (Figure \ref{lab_den_md3}b)
%causing a perturbation at the wind base. This perturbation propagates along the wind, causing the
%wind fan to oscillate as a whole, somewhat like a dancing flame in a breeze.
This effect is quite clearly presented in various panels (Figure \ref{lab_den_md3}c-f). It may also be noted that all the matter
%\textbf{coming out} 
coming out of the KD do not become a wind but sits above the KD.

\subsection{Angular momentum transport}
\label{sec_results_angular_momentum}
\begin{figure}
\centering
\includegraphics[width=80mm]{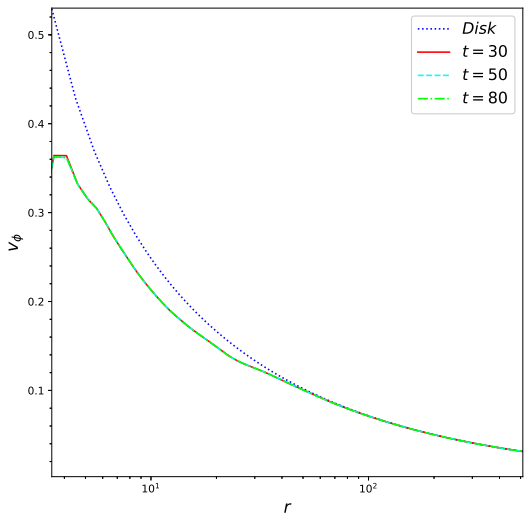}
\caption{%\textbf{Radial variation of $v_\phi$ \textbf{at a height $z=6$ from the equstorial plane,} 
Radial variation of $v_\phi$ at a height $z=6$ from the equatorial plane, for three run times $30$, 
$50$ and $80$ as shown in legends for $\dot{m} = 3$. The disc $v_\phi$ is plotted for comparison.}
\label{lab_v_phi}
\end{figure}

%\begin{figure*}
%\centering
%\includegraphics[width=170 mm]{fig5.eps}[!h] 
%\caption{Contours of $v_\phi$. Time elapsed is
%$t = 92$, for winds generated from accretion discs with (a) $\dot{m}=3$ and  (b) $\dot{m}=4$} 
%\label{lab_angmom}
%\end{figure*}
\begin{figure*}
  \begin{minipage}[c]{\textwidth}
    \begin{center}
        \includegraphics[width=150mm]{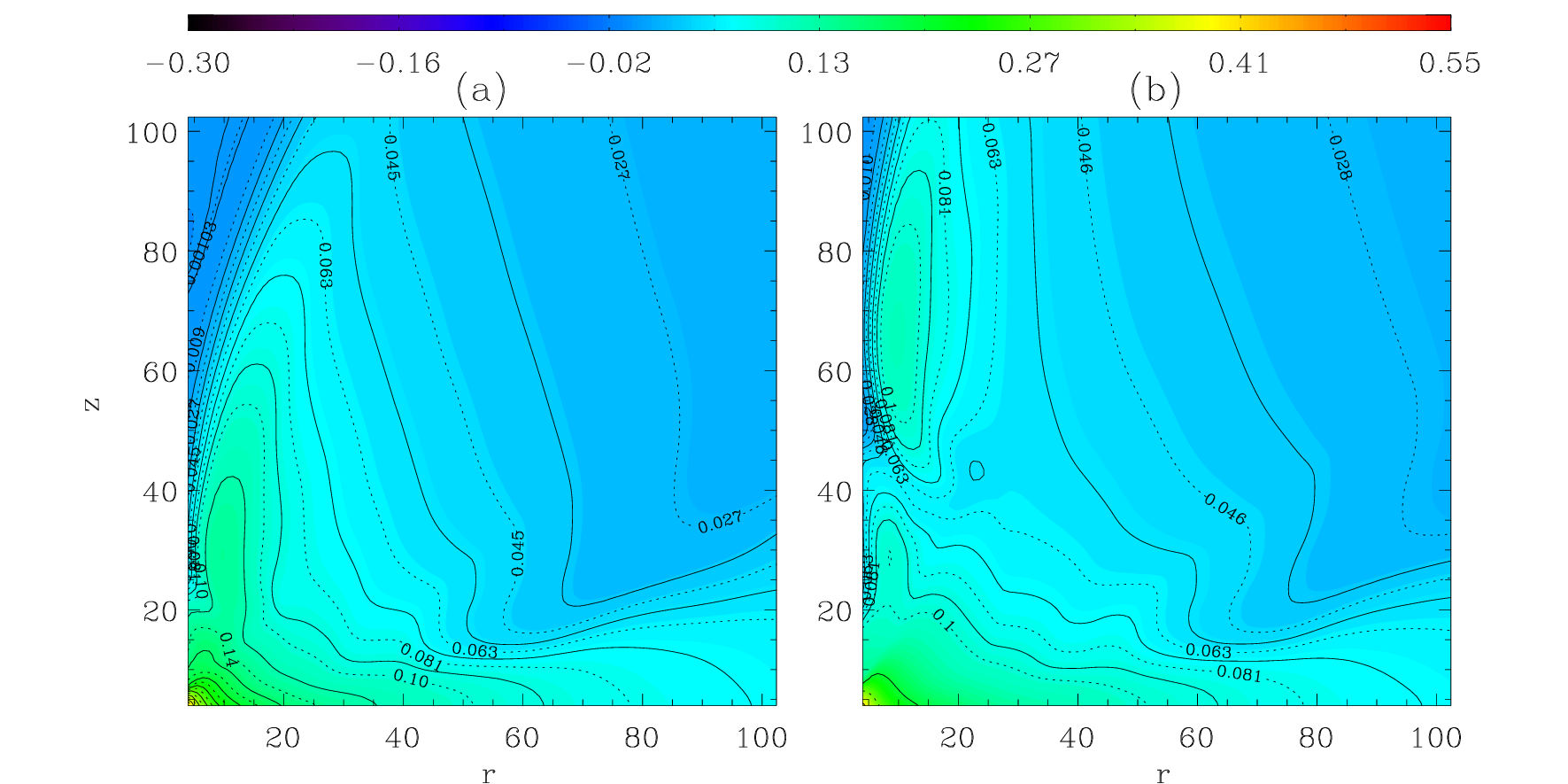} 
\caption{Contours of $v_\phi$. Time elapsed is $t = 92$, for winds generated from accretion discs with
(a) $\dot{m}=3$ and  (b) $\dot{m}=4$} 
	\label{lab_angmom}
    \end{center}
  \end{minipage}
\end{figure*}
Due to high rotational speed of the inner region of the disc, winds produced from this region
propagate with a fraction of
the rotational speed of the disc. Hence matter ejected from the KD carry a part of the disc
angular momentum along with them.
This can be seen as one of the ways
through which the disc removes its angular momentum. In Figure (\ref{lab_v_phi}), we plot
$v_{\phi}$ as a function of $r$ at a height of $z=6$ from the equatorial plane, measured in
three different %\textbf{
run%} 
times $t=30$ (solid), $50$ (dashed) and $80$
(dashed-dotted). %\textbf{
The disc $v_\phi$(dotted) is also included for comparison. It can be
seen that near the disc inner edge, almost $64\%$ of the azimuthal component of velocity
is effectively removed
by the wind which therefore would reduce angular momentum too.
%The angular momentum is effectively removed from the winds due to the drag term in
%the $\phi$ component of momentum equation.
All the terms with negative sign in the last term of
Equation(\ref{eq.momentum_phi}) resist rotation, thereby remove angular momentum (and
$v_\phi$) from the wind. %}
The rotational speed of the wind can be as
high as $0.3$ near the axis of symmetry but are much less than the disc rotational
velocity. Since the radiative moments are weak %\textbf{
above %} 
the outer part of the disc, the rotation
velocity %\textbf{
just %} 
above the disc is similar to that on the disc at the same $r$.
%Therefore, the rotational speed distribution decreases by an order magnitude as one moves outward
%by about $500 \rg$.
Since all the curves %\textbf{
(solid, dashed, dash-dotted) %} 
almost %\textbf{
overlap %} 
each other
%\textbf{
in Figure(\ref{lab_v_phi}),%},
we
conclude that the $v_\phi$ distribution close to the
disc is almost steady. %The time independence of $v_{\phi}$ close to the disc is obvious, as the disc rotation is steady.
Further, we plot the contours of $v_\phi$ of winds
generated from accretion discs with accretion rates $\dot{m}=3$ and $4$
%\textbf{
(Figures \ref{lab_angmom}a \& \ref{lab_angmom}b, respectively). %}
As the winds are stronger for higher accretion rates, the outflowing matter %\textbf{
driven by
radiation from a disc with %} 
higher $\dot{m}$,
matter with higher $v_\phi$ are injected. Winds from an accretion disc with $\dot{m}=4$, posses higher values of azimuthal
velocities
in a larger region above the accretion disc, compared to the winds from a disc of lower ${\dot m}$.
%\textbf{
It may be noted that, a fraction of outlfowing matter near the axis of symmetry fall back, and at some height above
the disc interacts with the outflowing wind and makes it to bend away. For KDs with higher $\dot{m}$ which are ejecting
fast matter and with higher rotation, traps these relatively higher rotating matter in the region where the inner boundary of the
wind bend away. And as the matter further move away, $v_\phi$ is reduced by radiation drag.
%and fraction of outflowing matter which falls back, perturbs the wind base, which causes the wind fan to dance like a flame in
%a breeze. This effect is accentuated for discs with higher $\dot m$. As the wind fan moves towards the axis, its poloidal speed
%$v_p$ may decrease and may create temporary blobs of matter with low $v_p$ but high $v_\phi$ as is seen near the axis and at height
%$z \sim 60$---$80$ in Figure (\ref{lab_angmom}b).
%Although the matter ejected from the disc has higher angular momentum, $v_{\phi}$ dominates above the disc.
%However, a portion of matter falls back to the base of the winds leading to a local disruption in their azimuthal rotation. So the
%rotation of the winds at the base decreases at later times and becomes much less as compared to the value mentioned above. This
%fall back of the matter is more prominant for higher $\dot{m}$ hence the disruption of $v_{\phi}$ at the base is more visible for
%$\dot{m}=4$ compared to $\dot{m}=3$.
%} 

\begin{figure*}
  \begin{minipage}[c]{\textwidth}
    \begin{center}
        \includegraphics[width=150mm]{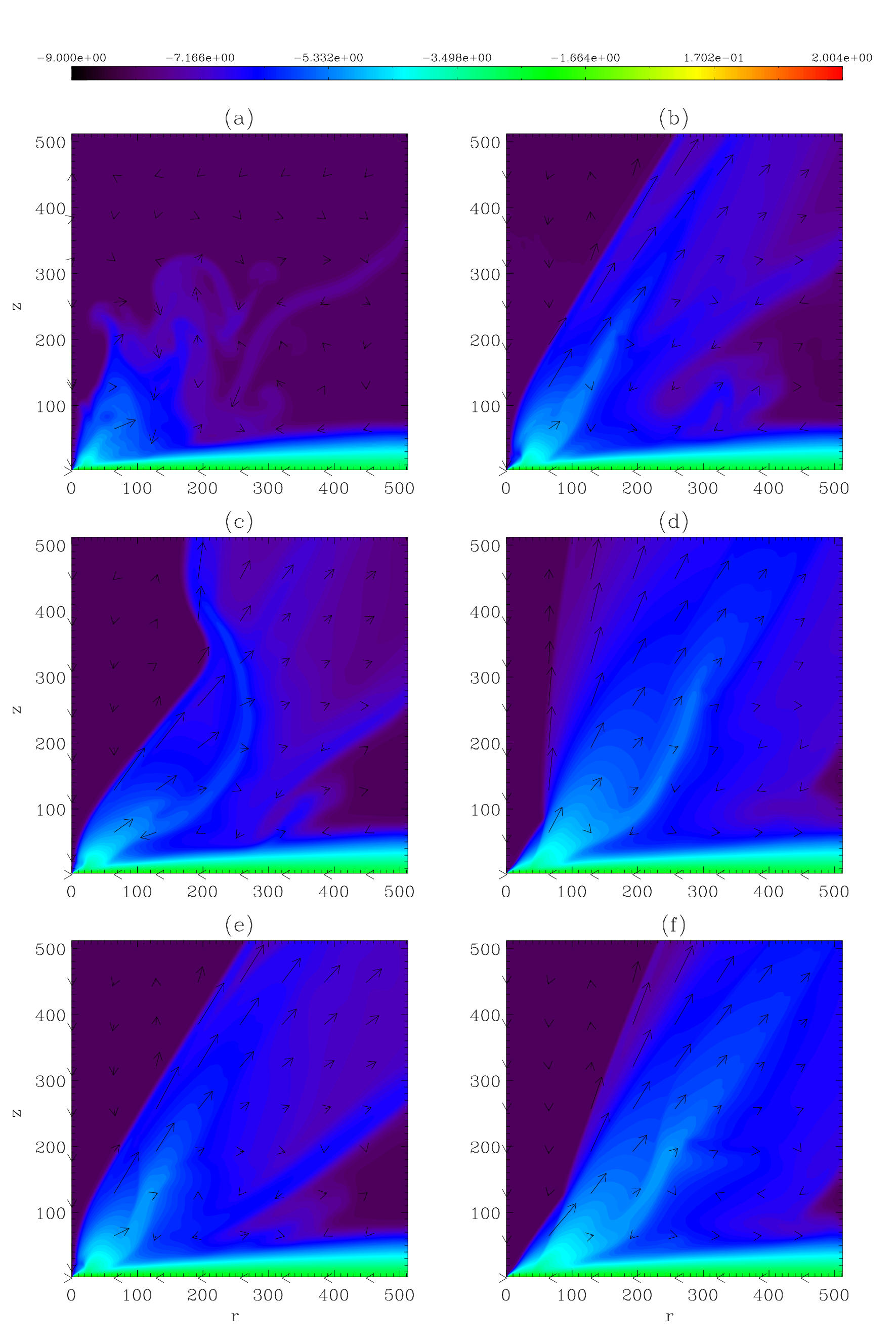} 
        \caption{Density distribution for $\dot{m}$ $\approx$ 1.3 (a \& b),  $\dot{m}$ $\approx$ 2.5 (c \& d) and
$\dot{m}$ $\approx$ 3 (e \& f), snapshots are at run time t = 72; frames in the left column are generated considering the
radiation drag and the right panel frames are without the drag effect. It is evident that for low mass accretion rate (here, $\dot{m}\approx 1.3$), the winds cannot be driven away due to the presence of the radiation drag.} 
	\label{lab_rad_drag1}
    \end{center}
  \end{minipage}
\end{figure*}

%\begin{figure*}
%\centering
%\includegraphics[width=150mm]{fig6.eps}
%\caption{Density distribution for $\dot{m}$ $\approx$ 2 (a \& b),  $\dot{m}$ $\approx$ 2.5 (c \& d) and
%$\dot{m}$ $\approx$ 3 (e \& f), snapshots are at run time t = 72: frames in the left column are generated considering the
%radiation drag and the right panel frames are without the drag effect. It is evident that for low mass accretion rate
%(here, $\dot{m}\approx 2$), the winds cannot be driven away due to the presence of the radiation drag.} 
%\label{lab_rad_drag1}
%\end{figure*}

\subsection{Effect of radiation drag}
In section (\ref{sec_equations}) the expression of the radiation term (equation \ref{radcomp.eq})
contains both positive and negative terms. The flux terms ($F_i$) are positive and therefore would accelerate the flow along its
direction.
However, the terms having radiation energy density and pressure components, appear with a negative sign and are also proportional
to various velocity components. The negative terms causes deceleration and would reduce relevant components of momentum density.
These negative terms are called radiative drag terms. 
For example,
the radial component of momentum density equation (\ref{eq.momentum_r}) will be increased by $F_r$, %\textbf{
but will
be reduced
provided any or all the terms containing $E$, $P_{rr}$, $P_{r\phi}$, $P_{rz}$ %}
are dominant. 
%As the fluid flows at supersonic speeds in a radiation field it
It may be noted that, the radiative drag terms are highly non-linear, for example,
$P_{\phi r}$ will couple with $v_r$ and hinder the growth of azimuthal momentum density ($\rho v_\phi$), but will also
couple with
$v_\phi$ and oppose the growth of $\rho v_r$ (refer to, equation(\ref{radcomp.eq}) for radiative terms and the equations of
motion \ref{eq.momentum_r}-\ref{eq.momentum_phi}).

%faces a resistance along its motion that slows it down. 
%This radiation is caused by radiation energy density and radiation pressure.
%As discussed in section (\ref{sec_results_angular_momentum}), the drag term in the $\phi$ component of
%the momentum balance equation (\ref{eq.momentum_phi}) takes away the angular momentum of the wind.
%Similarly, the drag term of the $z$ component of
%momentum balance equation (\ref{eq.momentum_z}) resists the flow speed of the wind in $z$ direction.
To show the impact of
radiation drag on the dynamics of the winds, we compare solutions with and without drag terms.
We plot the density contours and velocity field of wind solutions with drag terms in the left panels (a, c, e) of Figure
(\ref{lab_rad_drag1}), while solutions without
drag terms are plotted in the right panels (b, d, f) of the same figure.
The accretion rates for each pair of comparable panels are %\textbf{
${\dot m}=1.3$ %}
(Figures \ref{lab_rad_drag1}a \& b),
${\dot m}=2.5$
(Figures \ref{lab_rad_drag1}c \& d) and ${\dot m}=3$ (Figures \ref{lab_rad_drag1}e \& f).
%choose $\dot{m}=2$ and plot density contours for the wind solutions in
%Figures (\ref{lab_rad_drag1}a) and a corresponding density contour for the same wind solution without including the drag 
%terms in Eqs (\ref{eq.momentum_phi}, \ref{eq.momentum_r} and \ref{eq.momentum_z}) 
%in Figure (\ref{lab_rad_drag1}b).
All the plots obtained are at run time $t=72$. 
For lower ${\dot m}$ the wind is launched, but as it accelerates to higher velocities the drag terms suppress the wind.
However, in absence of radiation drag the wind freely
propagates outwards.
For a slightly higher
$\dot{m}(=2.5)$, a weaker wind is generated in presence of drag terms (Figure \ref{lab_rad_drag1}c) however, without drag terms
the wind is relatively stronger (Figure \ref{lab_rad_drag1}d). Similar
effect can be seen for $\dot{m}=3$ in which the wind in presence of drag terms (Figure \ref{lab_rad_drag1}e) is weaker than the
the one in absence of drag term (\ie  Figure \ref{lab_rad_drag1}f}). It may be noted, in absence of
radiation drag term, a lower luminosity disc will produce stronger winds than that above a luminous disc in which
drag terms are considered (compare Figures \ref{lab_rad_drag1}d \& \ref{lab_rad_drag1}e). Here Figure \ref{lab_rad_drag1}(e) is
identical
to Figure \ref{lab_den_md3}(d). Radiation driving ($F_r,~F_z$) is weaker above low luminosity accretion discs.
As the wind is launched from such a disc, it has low poloidal velocity ($v_p=\sqrt{v_r^2+v_z^2}$) to start with,
but high $v_\phi$. It means
radiation drag is not important
along $r,~z$ directions, but, $v_\phi$ being high, it boosts the drag terms in all the three direction.
Therefore, matter ejected from the disc will not flow out as wind but will be smothered down by the drag term.
Above a luminous disc, however, jets are strongly driven in $r,~z$ direction and can overcome the radiation drag 
due to high $v_\phi$.
At larger distances, where poloidal velocity increases,
drag becomes more effective and limits the terminal speed of the outflow.
This figure illustrates the effect of radiative drag.
%In the corresponding snapshot ($t=72$) winds without drag terms are significantly stronger. 
%So, the winds with smaller $\dot{m}$ (below $2.5$) are not capable to escape the domain because corresponding radiation field is too
%weak to drive them. They just fall back to the disc while radiation drag makes the winds relatively weaker for higher $\dot{m}$.

\subsection{Terminal speeds of the winds}
\begin{figure}
\centering
\includegraphics[width=8.5cm]{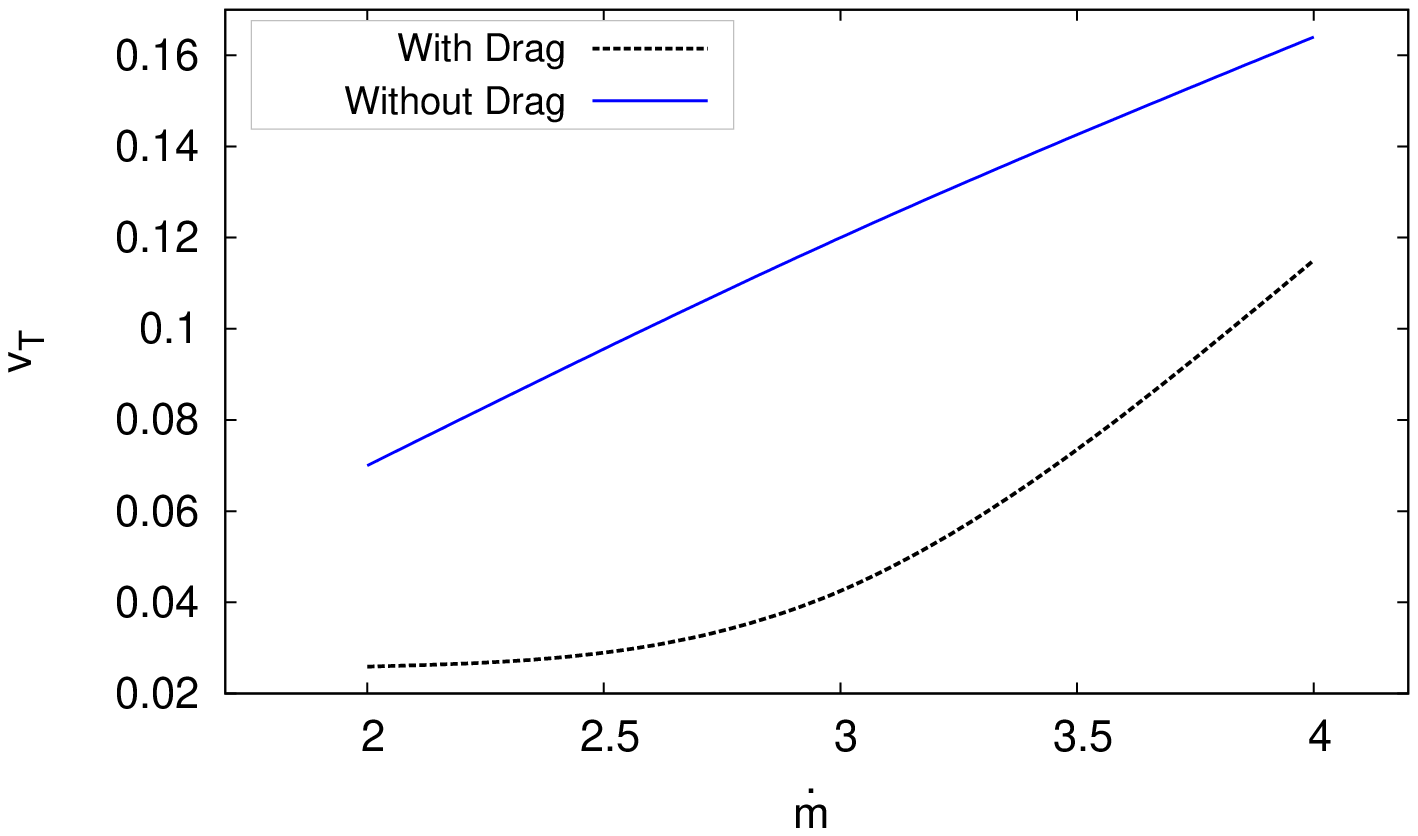}
\caption{Terminal speeds as a function of $\dot{m}$ for winds with radiation drag terms (solid curve) and without radiation drag
terms (dashed) at run time $t=100$.} 
\label{lab_rad_speeds}
\end{figure}

Once the winds leave the computational domain and escape to infinity, the maximum speeds they %\textbf{
acquire %}
while escaping is what we call
terminal speed ($\vt$).
%The terminal speed is defined as the maximum wind speed at the outer boundary of the domain at $t=\infty$ when the system
%relaxes to steady state. 
These are the speeds that correspond to the observed blue shift in the spectra from these sources. 
In this paper, we consider the the maximum speed at the outer boundary as the $\vt$ i.e.,
$\vt=v_p(r,512)$.
In Figure (\ref{lab_rad_speeds}), %\textbf{
we plot the terminal speeds as a function of $\dot{m}$ %}
at run time $t={100}$. The dashed curve corresponds to $\vt$ %\textbf{
when all the radiative terms are effective %}
(including drag terms). To show the effect of radiation drag,
we over plot corresponding terminal speeds 
without considering the radiation drag terms (solid). The winds are faster for higher accretion rates and safely reach mildly
relativistic values. In absence of radiation drag, the terminal speeds are overestimated by about an order of magnitude.

%{\bf
\subsection{Mass outflow properties}
\label{sec_outflow}
 \begin{figure}
\centering
\includegraphics[width=90mm]{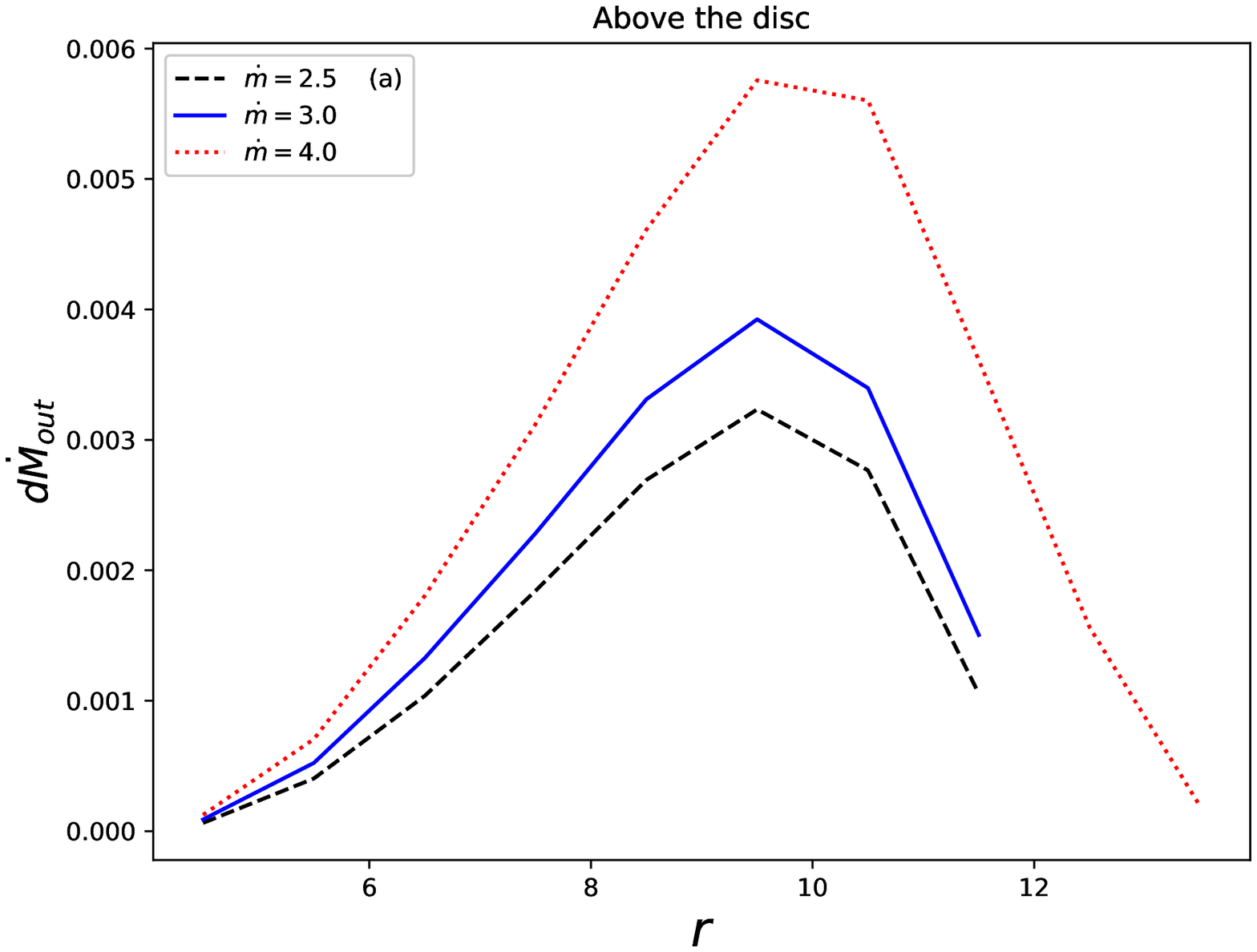}
\includegraphics[width=90mm]{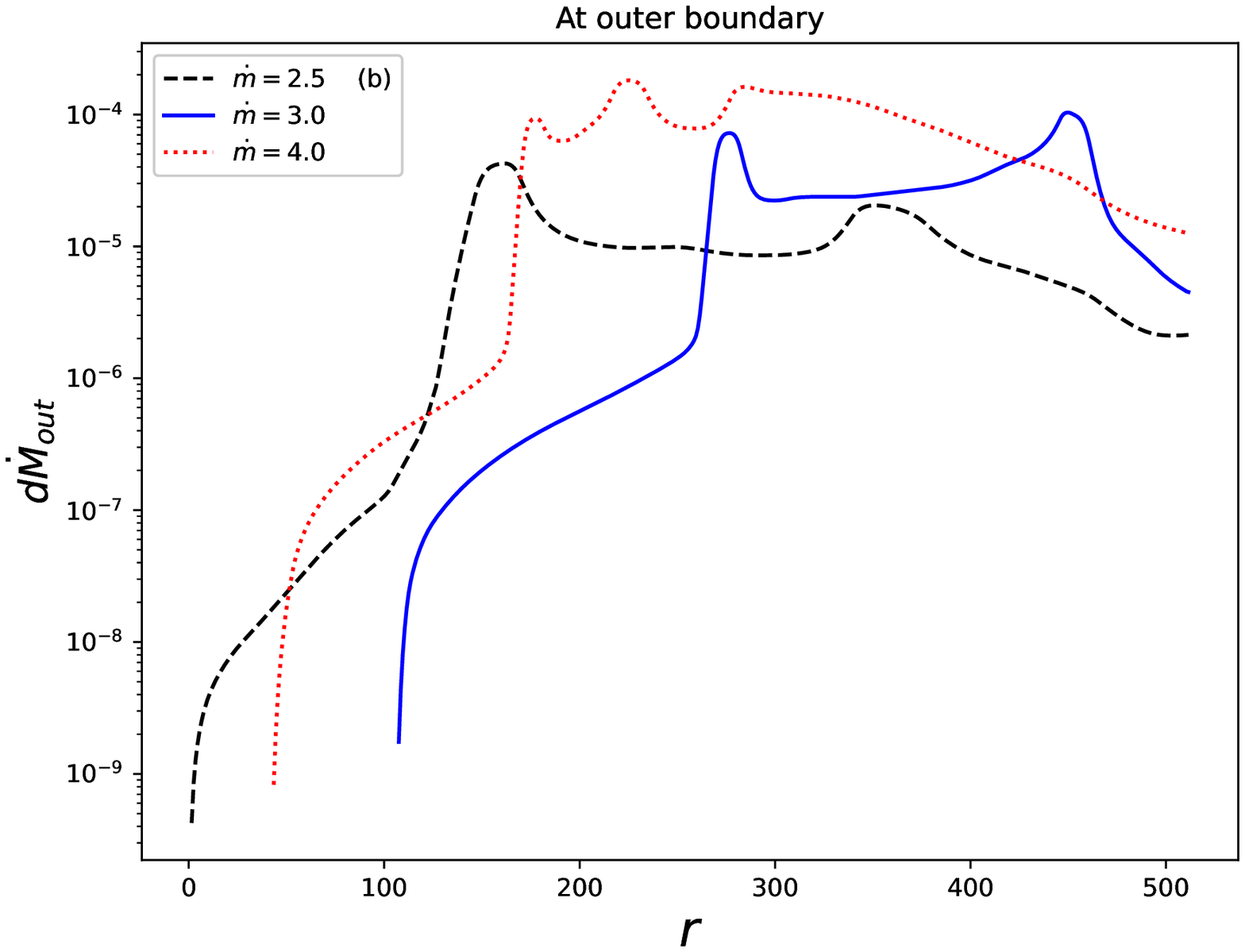}
\includegraphics[width=90mm]{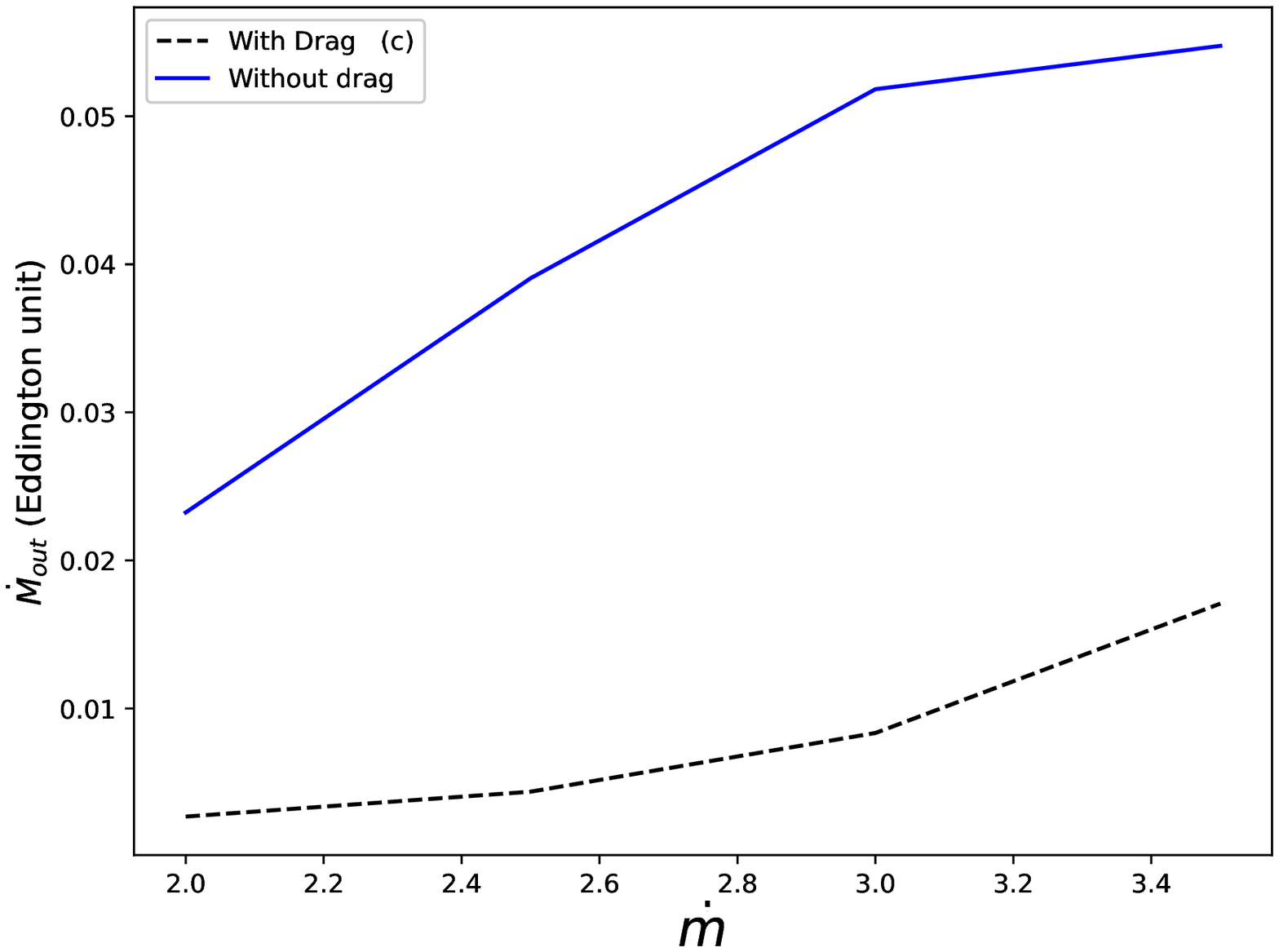}
\caption{Radial variation of mass outflow rate $d\dot{M}_{\rm out}(r)$, (a) just above the accretion disc and (b) at outer
$z$ boundary. (c) Variation of mass outflow ${\dot M}$ at the outer z boundary, w.r.t ${\dot m}$, with
and without considering the effect of radiation drag on the outflows. All the outflows are measured in Eddington unit, at run
time $t = 100$.}
\label{lab_mdout}
\end{figure} 

\begin{figure}
\centering
\includegraphics[width=90mm]{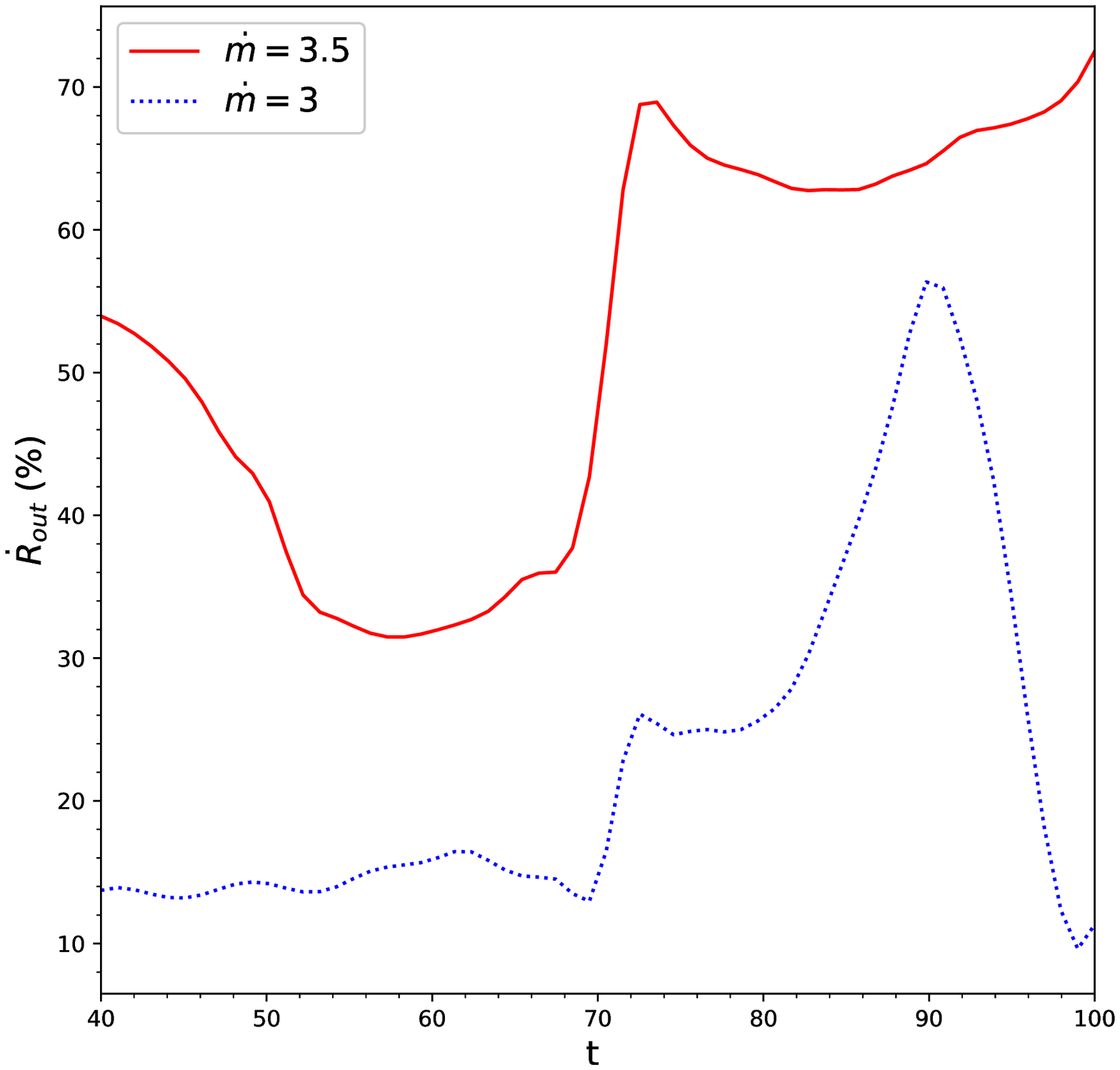}
\caption{The percentage of the total outflow which are transonic ${\dot R}_{\rm out}$ as a function of time,
for two accretion rates $\dot m=3.0$ (dotted, blue) and ${\dot m}=3.5$ (solid, red).}
\label{lab_transwind}
\end{figure} 
As the matter is ejected from the upper disc surface, the mass flux due to the emission can be represented in differential
form as
%The mass outflow rate in the fluid flow is estimated as (see ouyed,1997)
\begin{equation}
d\dot{M}_{\rm out} (r) =  2\pi r \rho v_z dr
\end{equation}
The total integrated value of $\dot{M}_{out}$ along radial direction can be written as
\begin{equation}
\dot{M}_{\rm out} = \int_{r_{i}}^{r_{o}} 2\pi r \rho v_z dr
\end{equation}

Here we have spatial resolution of $dr=1$. We calculate the radial variation of outflow at a certain height, z from the disc, as
well as the net integrated outflow along r at that specific z, for a particular time step. In Figure \ref{lab_mdout}(a), we plot
$d{\dot M}_{\rm out}$ (in Eddington units) with $r$, calculated just above the disc %(Essentially the mass outflow rate from the annular region $2\pi r dr$) 
for $\dot{m} = 2.5, 3$ \& $4$ at run time $t=100$, signifying the outflow from the launching point of the wind. Further, in
Figure \ref{lab_mdout}(b), we estimate the radial variation of the outflow rate at the outer $z$ boundary of the computational domain.
The outflow at the ejection base mostly comes out from the inner regions of the accretion disc and with time it gradually covers the
entire numerical domain diagonally leaving the domain through the outer boundaries. Further, the  outflow rates at the outer
boundary are 
significantly less than the outflow at the launching, indicating that all the matter ejected from the disc does not leave the
domain but only a fraction of it escapes. In Fig. \ref{lab_mdout}(c), we show the integrated outflow rates
${\dot M}_{\rm out}$ (black dashed)
from the disc calculated at outer $z$ boundary ($z=512$). These values are obtained by integrating the outflow rates along $r$,
at run time $t=100$. For comparison, we show the outflow rates without drag (blue solid) and observed that, as expected, the
radiation drag has significant effect in suppressing the matter ejection in form of the winds.
Furthermore, very small magnitudes of the outflow rates compared to the accretion rates justify our assumption that the disc can
remain mostly in steady state and is not affected by the matter ejection. In other words, the accretion rates remain time-independent.

The computational domain in this paper is just $512\times 512$, so it is intriguing to wonder what fraction of the computed outflow
will actually escape the gravity of the black hole. Since the wind is also a fluid, therefore if the wind is transonic
then it will definitely escape the the black hole gravity. In Fig. \ref{lab_transwind} we plot the percentage of
the calculated mass outflow rate at $z=512$ which are transonic or $v_T(r,512)/c_s(r,512) > 1$, and is measured as
$${\dot R}_{\rm out}=\frac{{\dot M}_{\rm out}(\mbox{trans})}{{\dot M}_{\rm out}(\mbox{total})},$$
where both ${\dot M}_{\rm out}(\mbox{trans})$ and ${\dot M}_{\rm out}(\mbox{total})$ are measured at $z=512$ or the upper
boundary of the computational box.
Figure \ref{lab_transwind} shows that of all the matter which is ejected from the computational domain,
only a fraction of it is transonic
and therefore actually can leave the gravitational attraction of the central black hole. For $\dot{m}=3$ (dotted, blue)
the transonic mass outflow rate is about 10\% of the total mass leaving the computational domain as winds. For $\dot{m}=3.5$
(solid, red) it varies between 30-70\% of the total outflow. We compute the wind flowing
out of the computational domain only through upper $z$ boundary, because mass-outflow rate through outer $r$
boundary is about an order of magnitude less compared to that through the upper $z$. Moreover, the mass flowing out through
the outer $r$-boundary is subsonic and would not contribute significantly in the net outflow rate. The wind outflow rate is also variable.
%}

\section{Conclusions}
\label{sec_conclusions}
In this paper, we have studied the generation mechanism and properties of the winds around black hole accretion discs. The winds are generated by a
Shakura - Sunyaev Keplerian accretion disc which is steady in nature and act as a source of wind and the radiation field which drives them. The radiation
field is controlled by $\dot{m}$. We computed all components of radiative moments numerically. The radiation field generated by the
steady KD are also
steady.  %\textbf{
It may be noted that, our assumption of optically thin nature of the medium above the KD is justified since the 
cumulative optical depth along the $z$-direction is much less than 1 (see appendix A). %}
Although we keep our analysis in non relativistic regime, we have used pseudo-Newtonian gravitational potential to take care of
strong gravity near
the black hole.

We show	 that the radiation pressure inside the disc along with the thermal pressure is able to push the matter out of the disc.
The winds are mostly generated
from inner region of the accretion disc ($r<30$). The matter emitted out not only carries matter with it but also removes
angular momentum from the disc. Highly rotating winds are driven by a combined effect of thermal pressure, radiation field and
centrifugal force and typically for $\dot{m}>1.8$ the winds escape to infinity. While for smaller accretion rates we
showed that the winds fall back
to the disc as they don't have sufficient radiation drive to push them to escape. One curious fact which these
simulations showed is that a part of the matter ejected does not become wind but may accumulate
above the disc.

We showed that the radiation drag limits the jet speed. In fact below
a certain luminosity, the wind is destroyed by the drag term. Only for a luminous disc, radiation can generate a wind
against gravity and its own drag. So radiation drag is a significant factor in determining the dynamical properties of the winds.
The $\phi$ 
component of the radiation drag is also capable to reduce the angular momentum of the wind.
%\textbf{
The work of \citet{Y18} is similar to ours, except that they considered outflows from a corona and they did not
consider radiation drag. These authors considered more luminous disc (up to 0.75 Eddington luminosity) while we considered
only up to 0.66 Eddington luminosity ($\equiv 4 {\dot M}_{\rm Edd}$). However, the maximum terminal speeds are somewhat similar
for luminous disc, although we predict a lower cutoff of disc luminosity to drive a wind from KD. %}

We analyzed the terminal properties of the winds and found that the terminal velocities of the disc winds are sub relativistic and
higher accretion rate
leads to higher magnitudes of the wind speed. The wind speeds are found to be mildly relativistic which is consistent with
observations. We show that if
radiation drag is ignored, the terminal speeds are overestimated significantly.% and %\textbf{
%may also be %}almost one order of magnitude higher.
%\textbf{
Detailed study of mass outflow rate shows that the mass loss from the disc is indeed a very small fraction of the disc mass
and hence we may conclude that the radiative property of the KD will not be significantly affected by radiatively driven
winds. Inclusion of radiation drag sufficiently suppresses the mass outflow rates at outer
boundary of our computational domain. %} 
So one needs to take care of
radiation drag effects while carrying out the analysis of radiation driving in the winds. 
It is a non relativistic study of the disc wind dynamics under impact of radiation field in Thomson scattering regime.
In upcoming works, we would examine the
role of Compton scattering in driving such winds. 

%\begin{figure}
%\centering
%\includegraphics[width=90mm]{ini0.eps}
%\caption{Normalized density distribution in the computational domain at t=0, the disc resides at the bottom.} 
%\label{Fig1}
%\end{figure} 

\section*{Acknowledgments} 
%\textbf{
The authors would like to thank the anonymous reviewer for insightful comments and suggestions that help us to improve the
manuscript. SR acknowledges the hospitality extended by ARIES during her many academic visits. MKV acknowledges his brief
postdoc tenure in ARIES where this work was initiated. %}

\section*{Data Availability}
The data underlying this article will be shared on reasonable request to the corresponding author.

\vspace{2mm}
%\textbf{
\appendix
\label{app:1}
\section{Calculation of optical depth of the radiation-driven-winds}
\begin{figure}
\centering
\includegraphics[width=90mm]{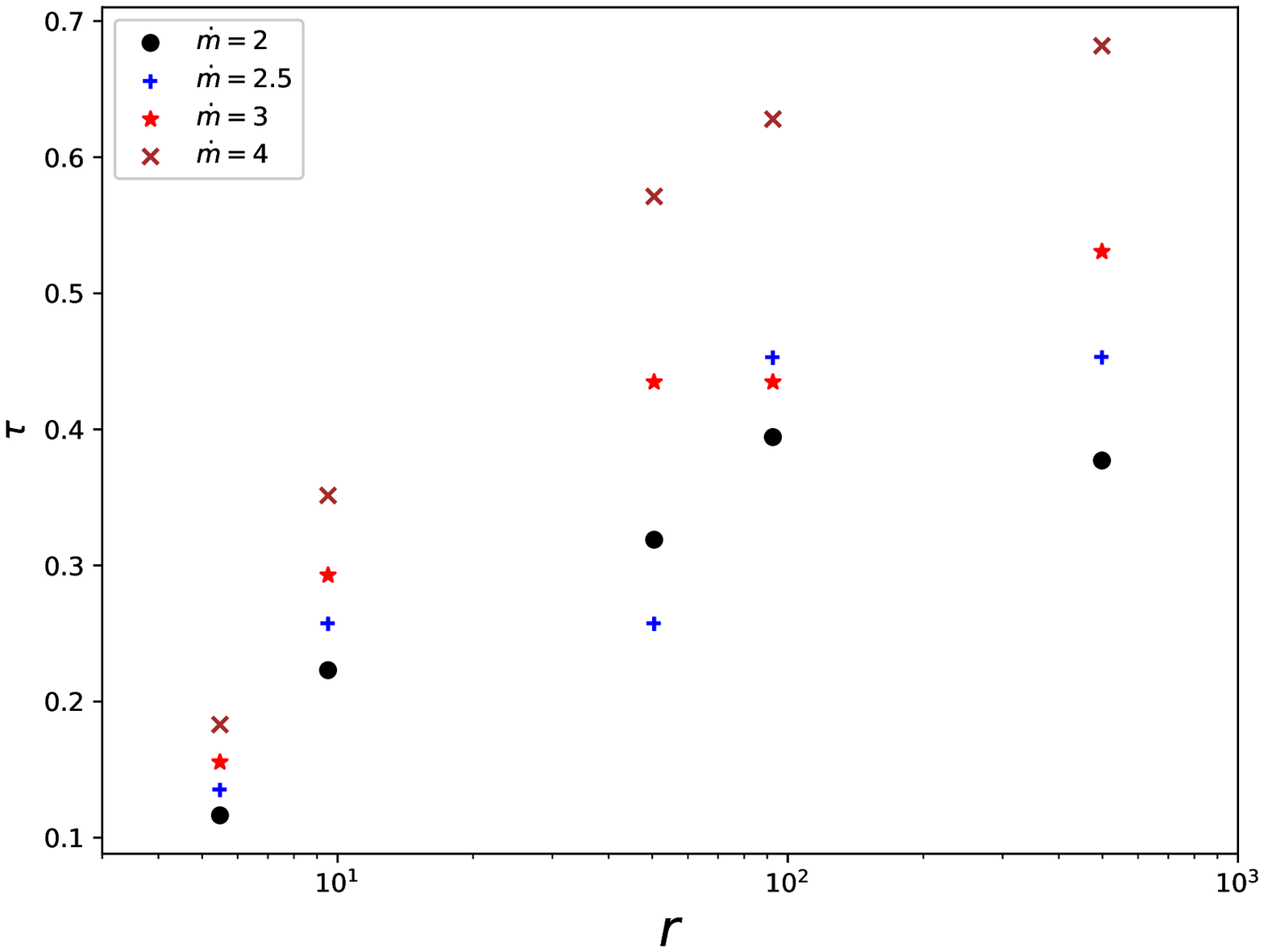}
\caption{{Cumulative optical depth $\tau$ for various values of $\dot{m}=2$ (Black circle),
{$\dot{m}=2.5$ (blue $+$), ${\dot m}=3$ (red star)} {and ${\dot m}=4$ (brown cross)}. The optical depth has been calculated
by integrating along $z$ at particular values of $r$.}}
\label{lab_optical_depth}
\end{figure}

The radiative moments in this paper are computed assuming that the medium above the disc is optically thin.
In this appendix, we calculate the optical depths above the accretion disc to check the justification to this assumption.
The optical depth is calculated by integrating the differential optical depth along $z$ direction, above the disc, in the
computational domain as 
\be 
\tau = \int_{z}\frac{\sigma_T \rho(z)}{m_p}dz %~~{\rm at}~~r=r_\tau
\ee 
It is calculated at a fixed radial distance $r$ from the black hole. In Figure (\ref{lab_optical_depth}) we plot the 
estimated optical depths above the disc for various values of $\dot{m}$ calculated at different  but fixed values of
$r$. Since, $\tau<1$ at every r
throughout the computational domain, the assumption of tenuous winds above the disc plane is justified. For high $\dot{m}$,
the optical depth
\textbf{reach up to $<0.7$} at the top of the computational domain at higher values of $r$, while it remains sufficiently low at smaller $r$
even for high $\dot{m}$.
It is worth mentioning that all the
radiative effects are mainly applicable to small $r$. At larger distances from black hole, the radiation field is not very
effective. %}

\end{document}